\documentclass[lettersize,journal]{IEEEtran}
\usepackage{amsmath,amsfonts}
\usepackage{algorithmic}
\usepackage{algorithm}
\usepackage{enumitem}
\usepackage{array}
\usepackage{textcomp}
\usepackage{stfloats}
\usepackage{url}
\usepackage{verbatim}
\usepackage{graphicx}
\usepackage{cite}
\usepackage{mdwmath}
\usepackage{booktabs}
\usepackage{mdwtab}
\usepackage{amsthm}
\usepackage{makecell}

\usepackage{xcolor}
\def\code#1{\texttt{#1}}
\newtheorem{definition}{Definition}
\usepackage{etoolbox}
\makeatletter
\patchcmd{\@begintheorem}{\textit}{\textbf}{}{}
\makeatother
\usepackage{amsmath,amsfonts}
\usepackage{algorithmic}
\usepackage{algorithm}
\usepackage{array}

\usepackage[font=normalsize,labelfont=sf,textfont=sf]{subfig}
\usepackage{graphicx}
\usepackage{cite}
\usepackage{comment}
\usepackage{balance}
\begin{document}

\title{SPGNN-API: A Transferable Graph Neural Network for Attack Paths Identification and Autonomous Mitigation}
\author{\IEEEauthorblockN{Houssem Jmal\IEEEauthorrefmark{1}\textsuperscript{\textsection},
Firas Ben Hmida\IEEEauthorrefmark{1}\textsuperscript{\textsection}, Nardine Basta\IEEEauthorrefmark{2}\textsuperscript{\textsection}, Muhammad Ikram\IEEEauthorrefmark{2}, Mohamed Ali Kaafar\IEEEauthorrefmark{2} and \\
Andy Walker\IEEEauthorrefmark{3}}\\
\IEEEauthorblockA{ \IEEEauthorrefmark{1}Ecole Polytechnique de Tunisie,\IEEEauthorrefmark{2}Macquarie University,
\IEEEauthorrefmark{3}Ditno INC.}}



\maketitle
\begingroup\renewcommand\thefootnote{\textsection}
\footnotetext{Authors contributed equally.}

\begin{abstract}
Attack paths are the potential chain of malicious activities an attacker performs to compromise network assets and acquire privileges through exploiting network vulnerabilities. Attack path analysis helps organizations to identify new/unknown chains of attack vectors exposing critical assets, as opposed to individual attack vectors in signature-based attack analysis. Timely identification of attack paths enables proactive mitigation of threats. Nevertheless, manual analysis of complex network configurations, vulnerabilities, and security events to identify attack paths is rarely feasible. This work proposes a novel transferable graph neural network-based model for shortest path identification. The shortest path, integrated with a novel holistic model for identifying potential network vulnerabilities interactions, is then utilized to detect network attack paths. Our framework automates the risk assessment of attack paths indicating the propensity of the paths to enable the compromise of highly-critical assets (e.g., databases). The proposed framework, named SPGNN-API, incorporates automated threat mitigation through a proactive timely tuning of the network firewall rules and Zero-Trust (ZT) policies to break critical attack paths and bolster cyber defenses. Our evaluation process is twofold; evaluating the performance of the shortest path identification and assessing the attack path detection accuracy. Our results show that SPGNN-API largely outperforms the baseline model for shortest path identification with an average accuracy $\geq$ 95\% and successfully detects 100\% of the potentially compromised assets, outperforming the attack graph baseline by 47\%.
\end{abstract}

\section{Introduction}
\IEEEPARstart{C}{yber} attacks have become not only more numerous and diverse but also more damaging and disruptive. New attack vectors and increasingly sophisticated threats are emerging every day. Attack paths, in general, are the potential chain of malicious activities an attacker performs to compromise assets and acquire network privileges through exploiting network vulnerabilities. Attack path analysis helps organizations identify previously unknown chains of attack vectors that could compromise critical network assets. 

Timely identification of attack paths enables proactive mitigation of threats before damage takes place. Nevertheless, manual processes cannot always provide the proactivity, fast response, or real-time mitigation required to deal with modern threats and threat actors, and constantly growing and dynamic network structure. An automated and efficient threat identification, characterization, and mitigation process is critical to every organization's cybersecurity infrastructure. 

The existing literature proposes various approaches based on attack graphs/attack trees that assess the dependencies between vulnerabilities and the potential impact of exploitation~\cite{byres, mcqueen, AWAN2016}. While these techniques provide a systematic perspective on potential threat scenarios, their effectiveness is constrained by their inability to dynamically adapt to changes in the network structure, thus requiring the re-evaluation of the entire process.

Several approaches based on deep learning (DL) have been proposed in the literature~\cite{DL_vul1, dl_vul2, dl_vul3} to address this issue. For such models, network structure information is not learned, unlike Graph Neural Networks (GNN), but rather provided as input to the DL models. Consequently, the structure-based input must be re-generated every time there is a change in the network structure. This can potentially necessitate the entire DL models to be retrained, causing additional computational overhead.

Another limitation of existing approaches is either being restricted to a set of predefined attacks~\cite{dl_vul2} or using a set of predefined rules to define the potential interplay between vulnerabilities~\cite{mulval2}. Given the rising complexity of cyber-attacks, a comprehensive approach rather needed. 

\noindent \textbf{Challenges.} There are three major challenges for attack path detection: (1) \textbf{Adaptiveness}: How to develop an automated and adaptive identification of attack paths given the dynamic nature of the network structure driven by trends such as remote users, bring-your-own devices, and cloud assets? (2) \textbf{Agility}: With attackers constantly finding new ways to exploit vulnerabilities, how to comprehensively identify the potential interplay between vulnerabilities without being bound to a pre-defined set of rules or attack scenarios? (3) \textbf{Efficiency}: How to efficiently characterize and rank the risks of attack paths, and autonomously triage the ones requiring prompt response without disrupting the network functionalities? 

\noindent \textbf{Our Work.} Considering these challenges, we devise ``Shortest Path Graph Neural Network-API'' ($\mathit{SPGNN-API}$), a framework offering an autonomous identification of potential attack paths and associated risks of compromising critical assets. It further incorporates proactive mitigation of high-risk paths. (1) To address the adaptiveness challenge, we develop a novel GNN approach for attack path identification. The inductive property of GNNs enables them to leverage feature information of graph elements to efficiently generate node embeddings for previously unseen data. Additionally, GNNs incorporate network structural information as learnable features. This renders GNN-based approaches self-adaptive to dynamic changes in the network structure. (2) To tackle the agility challenge, we assume that an attacker who has compromised an asset can exploit all the underlying vulnerabilities. 

We rely on the GNN efficiency of graph representation learning to learn all potential vulnerability interactions that could compromise critical assets based on the CVSS base score metrics~\cite{NVD}. (3) To address the efficiency challenge, we automate the risk analysis of attack paths to determine their likelihood of compromising critical assets, based on factors such as network configuration, assets' criticality, and the severity of the vulnerabilities~\cite{cvss2} in-path to the asset. We then develop autonomous mitigation of high-risk attack paths by automatically tuning the network ZT policies (See Section~\ref{sec:ZT}) to disrupt the paths without impacting the network functionalities.


In this work, we address a key limitation of existing GNNs that fail to capture the positional information of the nodes within the broader context of the graph structure~\cite{Spagan, distanceEncoding}. For instance, when two nodes share the same local neighborhood patterns but exist in different regions of the graph, their GNN representations will be identical. To address this, we introduce the $\mathit{SPGNN-API}$, which extends the Positional Graph Neural Network model~\cite{PGNN} to achieve a transferable model for computing shortest paths to a predefined set of nodes representing highly-critical network assets. 

\noindent \textbf{Evaluation.} We conduct a three-fold evaluation process: Firstly, we evaluate the performance of the $SPGNN$ shortest path calculation in a semi-supervised setting. Secondly, we assess the performance in a transfer-learning setting. Thirdly, we evaluate the accuracy of identifying critical attack paths. To carry out our evaluation, we use two synthetic network datasets, two real-world datasets obtained from middle-sized networks, and two widely used citation network datasets: Cora~\cite{cora} and Citeseer~\cite{CiteSeer}. We compare the GNN path identification performance with the state-of-the-art GNN path identification model $SPAGAN$~\cite{Spagan}. Additionally, we compare the performance of the $\mathit{SPGNN-API}$ with a state-of-the-art approach for attack path generation, $\mathit{MulVAL}$~\cite{mulval2}.

\noindent\textbf{Contributions.} In summary, our research contributions are: 
\begin{itemize}[noitemsep,topsep=0pt]
\item We develop a novel transferable GNN for shortest path calculation that relies exclusively on nodes' positional embedding, regardless of other features. The presented approach is able to transfer previous learning to new tasks, hence alleviating the lack of labeled data problems.
\item We propose a novel GNN-based approach for network vulnerability assessment and potential attack path identification that leverages the inductive ability of the GNNs to accommodate the dynamic nature of enterprise networks without requiring continuous retraining.
\item We demonstrate that, unlike traditional GNN, the performance of positional GNN models is enhanced by removing self-loops achieving an average improvement $\approx$ 5\% on our six datasets with a maximum of 9\%. 
\item We develop a novel comprehensive model for learning the propensity of vulnerabilities to contribute to attacks compromising critical assets based on the CVSS base metrics without being bound to specific attack signatures or a pre-defined set of rules for vulnerabilities interactions.
\item We formulate an autonomous risk characterization of the detected attack paths based on the network connectivity structure, asset configurations, criticality, and underlying vulnerabilities CVSS base score metrics. 
\item We automate the mitigation of high-risk attack paths that could potentially compromise critical assets by tuning the network's ZT policies to break the path without disrupting the network functionalities.
\item We evaluate our approach, the $\mathit{SPGNN-API}$, against two baseline models: the $SPAGAN$~\cite{Spagan} for GNN-based shortest paths detection and $\mathit{MulVAL}$~\cite{mulval2} for attack paths identification. Our results show that $\mathit{SPGNN-API}$ outperforms the baseline models, achieving an average accuracy of over 95\% for GNN shortest path identification. Moreover, our approach successfully identifies 47\% more potentially compromised assets that were not detected by the baseline model, $\mathit{MulVAL}$.
\end{itemize}

The rest of the paper is organized as follows: In Section~\ref{sec:relatedWork}, we survey the literature. In Section~\ref{sec:background}, we overview the ZT network architecture on which we base the attack paths risk assessment and mitigation. We further review different GNN architectures and limitations. Section~\ref{sec:design} details the design of our $\mathit{SPGNN-API}$ framework. We evaluate our model and present our results in Section~\ref{sec:evaluation}. Finally, Section~\ref{sec:conclusion} concludes our paper.

\section{Related Work}
\label{sec:relatedWork}
There exist commercial products such as \cite{paloalto, PRTG} to enable real-time network monitoring and connectivity assessment.  Similarly, configuration management and vulnerability scanning tools are accessible in the form of solutions like \cite{rapid7, nessus}. However, to the best of our knowledge, an integrated tool that amalgamates both functionalities remains absent in the landscape. SPGNN-API aims to harness network connectivity information to autonomously detect and mitigate potential chains of exploits directed at compromising critical assets.

This work has two major contributions; a novel GNN approach for shortest path identification and an autonomous framework for detecting and mitigating attack paths in dynamic and complex networks. To highlight the novelty of our work, in this section, we survey the literature and differentiate our contributions from previous studies related to network vulnerability assessment and attack graphs generation (Sec.~\ref{sec:attackgr}) and GNN-based distance encoding and shortest path identification (Sec.~\ref{sec:gnnsp}).

\subsection{Network Attack Graph and Vulnerability Assessment} \label{sec:attackgr}
We classify the existing approaches for vulnerability assessment into three main categories: traditional attack graphs/trees, ML/DL-based frameworks, and GNN-based approaches. 

\noindent{\bf Traditional attack graphs/trees vulnerabilities assessment frameworks.} 
This class of models examines the interplay between the network vulnerabilities and the extent to which attackers can exploit them, offering a structured representation of the sequence of events that can potentially lead to the compromise of network assets~\cite{byres, mcqueen, AWAN2016}. However, a major limitation of these models is their inability to adapt to dynamic changes in the network structure. Any modification to the network structure requires the regeneration of the attack graph.

\noindent{\bf Deep learning vulnerabilities assessment frameworks.} 
Previous studies have explored the use of deep learning-based (DL) approaches for vulnerability assessment and attack path detection~\cite{DL_vul1, dl_vul2, dl_vul3}. To identify potential attack paths in a network, information about the network structure and configurations is essential. However, in DL-based approaches, the network structure information is not learned, unlike GNN, and instead, provided as input to the DL model. Therefore, the structure-based input needs to be re-generated every time there is a change in the network structure, which may also require retraining the entire DL model. 

\noindent{\bf Graph neural network vulnerabilities assessment frameworks.} Recently, several approaches based on GNN have been proposed for cyber security tasks such as vulnerabilities detection~\cite{Gnnvulnerability2}, 
anomaly detection~\cite{Gnnanomaly}, malware detection~\cite{Gnnmalware} and intrusion detection~\cite{Gnnintrusion}. However, these approaches, in particular the vulnerability detection models, do not include any risk evaluation process that can help prioritize the detected threats for proactive mitigation. 

The presented approach addresses the limitations of existing work by enabling the identification of the potential based on the CVSS base metrics without being bound to specific attack signatures or a pre-defined set of rules for vulnerability interactions. By relying on GNN for attack path identification, it inherently accommodates the dynamic nature of enterprise networks without requiring continuous retraining.

\subsection{GNN Shortest Path Identification}\label{sec:gnnsp}

The goal of graph representation learning is to create representation vectors of graphs that can precisely capture their structure and features. This is particularly important because the expressive power and accuracy of the learned embedding vectors impact the performance of downstream tasks such as node classification and link prediction.

However, the existing GNN architectures have limited capability for capturing the position/location of a given node relative to other nodes in the graph~\cite{GNNWEAK1} (See Sec.~\ref{sec:expressive}). GNN iteratively updates the representation of each node by aggregating representations of its neighbors. Many nodes may share a similar neighborhood structure, and thus, the GNN produces the same representation for them although the nodes may be located at different locations in the graph.

Several recent works have addressed this limitation of GNNs. Although some of these approaches have been successful, we present the first GNN-based method that is transferable and can accurately calculate shortest paths using only distance information, without relying on other node or edge features.

For instance, in~\cite{distanceEncoding}, the authors propose a general class of structure-related features called distance encoding, which captures the distance between the node set whose representation is to be learned and each node in the graph. These features are either used as extra node attributes or as controllers of message aggregation in GNNs.

The Positional Graph Neural Network (P-GNN)~\cite{PGNN} approach randomly samples sets of anchor nodes. This method then proceeds to acquire a non-linear vector through a distance-weighted aggregation scheme over the selected anchor sets. This vector represents the distance between a given node and each of the anchor sets. It's important to note that the PGNN model focuses on learning a representation of the shortest path but does not possess the ability to predict the actual shortest path distance. Additionally, its performance in pairwise node classification is found to be somewhat limited, with an accuracy rating below 65\% observed on two of the three evaluation datasets. Notably, these datasets are characterized by a lack of node features or limited feature availability, as highlighted in \cite{PGNN}.

Another approach, $SPAGAN$~\cite{Spagan}, conducts paths-based attention in node-level aggregation to compute the shortest path between a center node and its higher-order neighbors. $SPAGAN$, therefore, allows more effective aggregation of information from distant neighbors into the center node. This approach relies on node features to predict the shortest path distance, hence yields inaccurate results in the event only limited features of graph nodes are available.

To address the limitations of existing approaches the presented GNN for shortest path length values calculation relies exclusively on nodes' positional embedding, regardless of other features. The presented approach is able to transfer previous learning to new tasks, hence alleviating the lack of labeled data problems.



\section{Background}
\label{sec:background}
In this section, we overview the Zero-Trust Architecture (ZTA) and related policies' governance and compliance on which we base the risk assessment, triage, and mitigation of the detected attack paths (Sec.~\ref{sec:ZT},~\ref{sec:gov}). As the proposed framework relies on shortest paths calculation to identify attack paths, we briefly explain the shortest path identification problem (Sec.~\ref{sec:spc}) and discuss the processing of graph data with GNNs (Sec.~\ref{sec:Ognn}). We highlight the limitations of existing GNN architectures (Sec.~\ref{sec:expressive}) that have motivated our novel GNN-based model for shortest path identification.
\subsection{Zero-Trust Architecture}
\label{sec:ZT}
ZT is a comprehensive approach to secure corporate or enterprise resources and data, including identity, credentials, access management, hosting environments, and interconnecting infrastructure. ZTA can be enacted in various ways for workflows. 

For instance, micro-segmentation~\cite{ACSC} enforces ZTA by creating secure zones in cloud and data-center environments, isolating and securing different application segments independently. It further generates dynamic access network-layer control policies that limit network and application flows between micro-segments based on the characteristics and risk appetite of the underlying network's assets.

Micro-segmentation is implemented via a distributed virtual firewall that regulates access based on network-layer security policies for each micro-segment. By limiting access to only what is necessary, micro-segmentation helps to prevent the spread of attacks within a network. The ZT micro-segmentation policies are defined as: 
\begin{definition}\label{def:zt}
ZT policies refer to network layer policies that the micro-segmentation distributed firewalls enforce to control the internal communications of the network. These policies follow the format: \code{< Source Micro-Segment IP Range > < Destination Micro-Segment IP Range > < Protocol > < Port Range >}.
\end{definition}

\subsection{Governance and Compliance}
\label{sec:gov}
The visibility of the network micro-segments underlying assets' characteristics and criticality is crucial for the optimal management of network communication policies. To achieve this purpose, a semantic-aware tier, called ``governance'', is used with the ZT policies to ensure their compliance with the best practices for communication between the network assets~\cite{noms2022}. The governance tier uses semantic tags (e.g. Database, Web Server, etc.) to perform a risk-aware classification of the micro-segments and underlying assets based on the criticality of the data stored transported, or processed by the micro-segment assets and their accessibility~\cite{assetValuation}. 

In this work, we consider eight criticality levels for classifying the network micro-segments as detailed in Table~\ref{tab:crit}. This table is generated following the study in~\cite{assetValuation} in conjunction with guidance from the security team administrators of the two enterprises contributing to this study. It is worth mentioning that the governance rules are generated following the best network communication practices. They are tuned per organization based on the network structure and business processes. A governance rule is defined as follows:
\begin{definition}
 A governance rule represents the best practice of who/what communicates to the different network assets. It relies on the micro-segments assigned tags to assess the communications enabled through the network ZT policies. A governance rule has the following format:
\code{< Source Tag > < Destination Tag > < Service Tag >}. 
\end{definition}

The Governance module assesses the compliance of each ZT policy with the respective governance rule. Consider $\mathit{P}$ to be the set of governance rules. Governance-compliant connections, denotes by $\mathit{CC}$, are defined as follows:
\begin{definition}
Compliant connections are communications allowed by the ZT policies that comply with the defined governance rules. Let $\mathit{CC}$ denote the set of compliant edges (connections enabled by the ZT policies) where $\mathit{CC \subseteq \{ E \;| \; (tag(x), tag(y),s) \in P\}}$ and $\mathit{tag(v)}$ be a function to identify the governance tag assigned to vertex $\mathit{v \in V}$. 
\end{definition}

For instance, the ZT policy \code{< Human-Resources Web Server IP Address > < Human-Resources Application Server IP Address > < TCP > < 443 > }is compliant with the governance rule \code{< Web Server > < Application Server > < Secure Web >}. Hence, all communications enabled through the above ZT policy are marked safe. 

Similarly, we denote by $\mathit{NC}$ the set of non-compliant edges. In a network setting, {\it compliant} connections are usually considered trusted as per the governance policies. The criticality of the non-compliant connections amongst the assets is a function of the trust rating of its incident vertices i.e., assets. 

\begin{table}[h]
\centering

\begin{tabular}{|l|l|}
\hline
\textbf{Level} & \textbf{Description}\\
\hline
0 & UnTagged/unknown\\
1 & Untrusted and external/public e.g internet 0.0.0.0/0\\
2 & Trusted external e.g vendor\\
3 & Internet facing\\
4 & Untrusted and internal e.g users\\
5 & Internal and connecting to untrusted internal e.g web servers\\
6 & Internal and connecting to data or non-critical data\\
7 & Critical data\\

\hline
\end{tabular}
\caption{Assets criticality levels and associated description.}
\label{tab:crit}
\vspace{-0.2cm}
\end{table}

In this work, we are mostly concerned with attack paths potentially compromising highly-critical assets. In particular, the ones incorporating non-compliant connections which imply a relatively higher risk of being exploited. In this context, we define highly-critical assets as follows:
\begin{definition}
Highly-critical assets are network resources that are considered valuable due to the sensitivity of the data they host (e.g. databases). Let $\mathit{V_{critical}}$ denote a set of nodes with maximum criticality. Formally, $\mathit{V_{critical}= \{ v \;| \; v \in V \; \land \; c_{v} \;= \; 7 \}}$ where $\mathit{c_{v}}$ is the criticality rating of node $\mathit{v}$ implied by the assigned governance tag. 
\end{definition}

\subsection{Shortest Path Identification} \label{sec:spc}
Shortest path (SP) algorithms (e.g. Bellman-Ford, Dijkstra's) are designed to find a path between two given vertices in a graph such that the total sum of the weights of the edges is minimum. Our proposed framework relies on shortest paths calculation to identify the eminent worst-case scenario for potential cyber-attacks compromising highly-critical assets \cite{spatt,spatt2} and to implement the appropriate mitigation accordingly. In this context, we define a critical attack path as follows~\cite{Apath}: 
\begin{definition} \label{def:attack_path}
An attack path is a succinct representation of the sequence of connections (enabled by ZT policies) through vulnerable assets that an attacker needs to exploit to eventually compromise a highly-critical asset.
\end{definition}

The time complexity of shortest path (SP) algorithms on a directed graph can be bounded as a function of the number of edges and vertices by $\mathit{O(VE)}$~\cite{SP_complexity}. However, the complexity of SP algorithms can be improved by using GNNs to approximate the distance between nodes in a graph. After training a neural network, the time complexity of finding the distance between nodes during the inference phase is constant, denoted by $\mathit{(O(1))}$.



\subsection{Processing Graph Data with GNNs} \label{sec:Ognn}
The goal of graph representation learning is to generate graph representation vectors that capture the structure and features of graphs accurately. Classical approaches to learning low dimensional graph representations~\cite{deepwalk,skipgram} are inherently transductive. They make predictions on nodes in a single, fixed graph (e.g. using matrix-factorization-based objectives) and do not naturally generalize to unseen graph elements.

Graph Neural Networks (GNNs)~\cite{semiKipf, kipf2} are categories of artificial neural networks for processing data represented as graphs. 
Instead of training individual embeddings for each node, GNNs \emph{learn} a function that generates embeddings by sampling and aggregating features from a node's local neighborhood to efficiently generate node embeddings for previously unseen data. This inductive approach to generating node embeddings is essential for evolving graphs and networks constantly encountering unseen nodes.

GNNs broadly follow a recursive neighborhood aggregation (or message passing) where each round of neighborhood aggregation is a hidden layer $\mathit{l}$ in the GNN. Let $\mathit{G=(V,E)}$ denote a directed graph with nodes $\mathit{V}$ and edges $\mathit{E}$. Let $\mathit{N(v)}$ be the neighborhood of a node $\mathit{v}$ where $\mathit{N(v)=\left\{ u \in V \; | \; (v,u) \in E\right\}}$. For each layer, or each message passing iteration, a node $\mathit{v}$ aggregates information from its sampled neighbors $\mathcal{N}$ $\left(v\right)$ as described in Equation~\ref{eq:agg1}.
\begin{equation} \label{eq:agg1}
h_{v}^{l}= \sigma\left(M^{l} \cdot \Lambda\left(\left\{h_{v}^{l-1}\right\} \cup\left\{w_{e}{h}_{u}^{l-1}, \forall u \in \mathcal{N}(v)\right\}\right)\right)
\end{equation}

The aggregated information is computed using a differentiable function $\Lambda$ and a non-linear activation function $\sigma$. $\mathit{w_{e}}$ is the edge feature vector from node $v$ to node $u$. The set of weight matrices $\mathit{M^{l}, \forall l \in\{1, \ldots, L\}}$ are used to propagate information between layers. After undergoing $k$ rounds of aggregation, a node is represented by its transformed feature vector, which encapsulates the structural information of the node's k-hop neighborhood as described in~\cite{GNNIntro}. 


\subsection{GNNs Expressive Power} \label{sec:expressive}

The success of neural networks is based on their strong expressive power to approximate complex non-linear mappings from features to predictions. GNNs learn to represent nodes' structure-aware embeddings in a graph by aggregating information from their $k$-hop neighboring nodes. However, GNNs have limitations in representing a node's position within the broader graph structure~\cite{distanceEncoding}. For instance, two nodes that have topologically identical or isomorphic local neighborhood structures and share attributes, but are in different parts of the graph, will have identical embeddings.

The bounds of the expressive power of GNNs are defined by the 1-Weisfeiler-Lehman (WL) isomorphism test~\cite{GNNWEAK1} 
In other words, GNNs have limited expressive power as they yield identical vector representations for subgraph structures that the 1-WL test cannot distinguish, which may be very different~\cite{PGNN,distanceEncoding}. 

\section{Proposed Framework SPGNN-API}
\label{sec:design} 

In this section, we present our proposed framework that aims to achieve end-to-end autonomous identification, risk assessment, and proactive mitigation of potential network attack paths. To identify internal network connections and associated accessibility policies we rely on the Zero-Trust policies that govern the communication between the microsegments at the network layer. As depicted in Figure~\ref{fig:design}, the $\mathit{SPGNN-API}$ consists of five modules: (a) Network micro-segmentation, (b) governance and compliance, (c) network data pre-processing, (d) GNN-based calculation of shortest paths to critical assets, and (e) risk triage and proactive mitigation. We elaborate on these modules in the following subsections.
\begin{figure*}
    \centering
    \includegraphics[width=\textwidth]{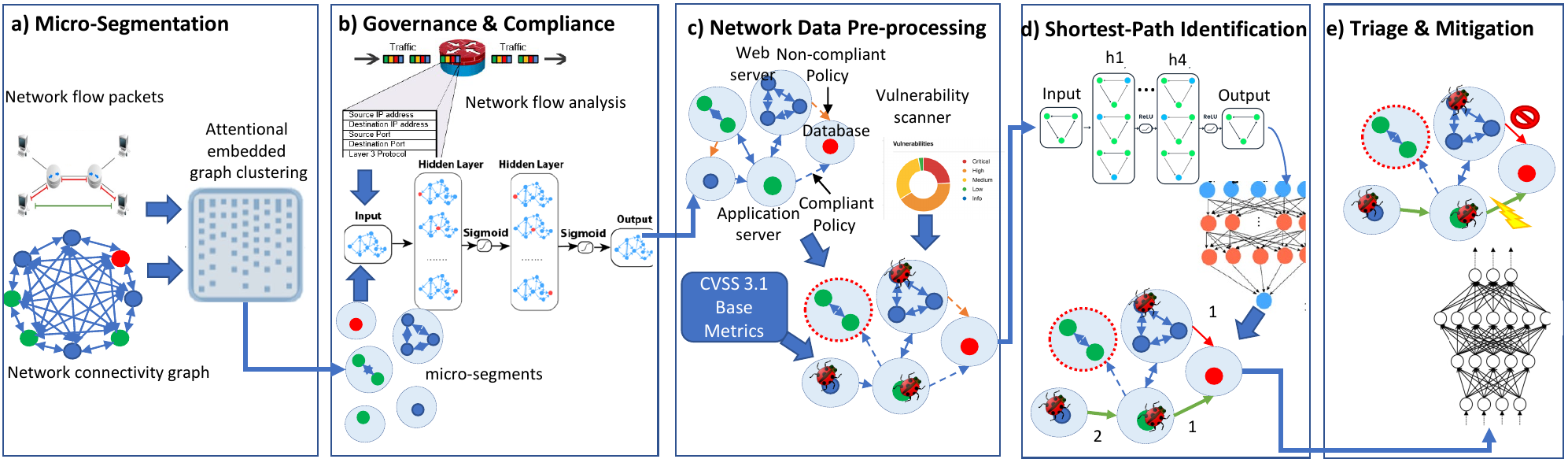}
    \caption{\small $\mathit{SPGNN-API}$ framework architecture where Sub-figure a) illustrates the micro-segmentation process through attentional embedded graph clustering of the network based on layer 2 and 3 flow packets header analysis and the network connectivity graph. This process is followed by a GNN-based model for generating the ZT policies governing the communication between the micro-segments as detailed in Sub-figure b). Sub-figure c) describes the network data pre-processing stage to illuminate the edges that cannot be part of an attack path. The updated graph is then used to identify the shortest paths to highly-critical assets as illustrated in sub-figure d). Finally, edges are classified as either safe, compliant critical, or non-compliant critical. The ZT policies are then tuned to block the latter class of edges.
    }
    \label{fig:design}
        \vspace{-0.25cm}
\end{figure*}
\subsection{Micro-Segmentation}
First, we represent a given network as a directed connectivity graph. Let $\mathit{C(V,E,S)}$ be a labeled, directed graph that represents the network's connectivity, where $\mathit{V}$ is the set of graph vertices representing the network assets (servers and cloud resources). The set of graph-directed edges $\mathit{E}$ indicates the connected vertices' communication using the service identified through the edge label $\mathit{s \in S}$. Here $\mathit{S}$ denotes the set of network services that are defined by a protocol and port range and $\mathit{E \subseteq \{ (v,u,s) \;| \; (v,u) \in V^2 \land x \neq y  \land s \in S\}}$. 

We derive the set of feature vectors characterizing the graph vertices (network assets) and edges (incident assets communication) from layers 3 and 4 network flow packet headers. This includes features such as frequently used ports and protocols, frequent destinations, and flow volume. Our approach assumes that assets within the same micro-segment exhibit similar communication patterns.

To automatically identify the network micro-segments, we use attentional embedded graph clustering~\cite{AGL}, a deep embedded clustering based on graph attentional auto-encoder. The clustering algorithm aims at partitioning the connectivity graph $\mathit{C=(V,E,S)}$ into $\mathit{k}$ sub-graphs representing the network micro-segments. It learns the hidden representations of each network asset, by attending to its neighbors, to combine the features' values with the graph structure in the latent representation. We stack two graph attention layers to encode both the structure and the node attributes into a hidden representation. 

\subsection{Governance and Compliance}
Each micro-segment is assigned a ``governance'' tag implying its underlying assets' criticality and risk appetite. For instance, a \textit{web server} asset criticality is lower than a \textit{database}. To automate the assignment of tags, we further assess the network flows in terms of communication patterns and frequently used ports and protocols to identify the dominating service(s) used by each micro-segment's underlying assets. For instance, a micro-segment mostly using TCP 80 for communication is most likely a web server. 
The detailed process of application profile assignment and the handling of dynamic ports is beyond the scope of this paper. 

We then automate the generation of the ZT policies to govern the communication between the micro-segments at the network layer. We first identify all attempted communications in the network and automatically generate ZT policies to enable all these communications. We compare the generated policies with the governance rules and highlight the non-compliant policies. We further assess the risk imposed by the non-compliant connections based on the criticality of the incident edges and the network topology. We then formulate a GNN model for tuning the ZT policies to reduce the risks without disrupting the network functionalities. The details of this process are beyond the scope of this paper.

\subsection{Network Data Pre-processing}\label{sec:attack}
$\mathit{SPGNN-API}$ relies on shortest paths calculation to predict imminent attack paths. We aim to pre-process the network connectivity graph by identifying edges that can potentially contribute to attack paths and filter out the edges that cannot be exploited by an attacker. This pre-processing stage ensures that all calculated shortest paths do represent attack paths.

An essential step toward the identification of an attack path is locating network vulnerabilities and assessing their severity which directly impacts the risk imposed by potential attacks exploiting these vulnerabilities. To locate the network vulnerabilities, we utilize a port scanner (e.g. Nessus). We then rely on the NIST Common Vulnerability Scoring System (CVSS) base metrics~\cite{cvss2} to identify the features and severity of the detected vulnerabilities.

We identify edges potentially contributing to critical attack paths following an exclusion methodology. We filter out edges that cannot be exploited by attackers based on a pre-defined set of criteria. This set of criteria does not define specific vulnerability interactions and ways of exploiting these vulnerabilities. They rather highlight the propensity of exploiting the vulnerabilities to eventually compromise critical assets.

\noindent\textbf{Edges exclusion criteria: }Graph edges are excluded if they don't meet the following criteria: (1) The edge source node needs to have a vulnerability with CVSS base metric ``scope'' set to ``changed''. This implies that the vulnerability can impact resources in components beyond its security scope. Hence, being exploited, it enables the attacker to move further in the network and potentially reach a highly-critical asset. (2) The edge source node needs to have a vulnerability with CVSS overall base score metric ``High'' or ``Critical''. This implies the potential criticality of the attack step. (3) All edges with highly-critical asset destinations are considered.

A major strength of our proposed approach is that it does not restrict the detection of potential attacks to a predefined set of vulnerability interactions. Instead, we assume that once an attacker gains access to an asset, they can exploit any underlying vulnerability without any specific prerequisites such as access rights or user privilege. This assumption is based on the constantly evolving nature of attacks and the ability of attackers to discover new ways of exploiting vulnerabilities. Consequently, we do not track an end-to-end sequence of attack steps as there might be infinite alternatives. Instead, we identify the propensity of an edge being involved in an attack by determining if there exists a (shortest) path from that edge to a highly-critical asset going through vulnerable nodes.

This comprehensive approach to representing vulnerability interactions is not feasible for traditional attack path detection models due to the time complexity of generating attack trees, where the size of the graph is a function of the potential vulnerabilities' interactions~\cite{mulval2}. However, our presented approach, which is based on the P-GNN, overcomes this issue with a time complexity of $\mathit{O(nlog^2n)}$, where $n$ is the number of assets in the network. Accordingly, the size of the graph is irrelevant to the number of vulnerabilities and their potential interactions.

\subsection{GNN Model for Shortest Paths Identification}
We formulate and develop a transferable GNN model for shortest path identification. Our approach involves identifying the shortest paths to a predefined set of nodes representing highly-critical assets in a network. By identifying the shortest path representing the minimum set of exploits an attacker would need to compromise highly-critical assets, we account for the worst-case scenario for potential attacks.

%

%
Our framework is based on the Position Graph Neural Network (P-GNN) model that randomly samples sets of anchor nodes. It then learns a non-linear vector of distance-weighted aggregation scheme over the anchor sets that represents the distance between a given node and each of the anchor sets~\cite{PGNN}. 


To enhance the P-GNN architecture; firstly, we recover the actual shortest path distance from the node embeddings through a transferable GNN model. Secondly, we identify the shortest path length to a predefined set of nodes representing high-criticality assets rather than a randomly distributed set of anchors. Thirdly, we update the message function to only consider the position information for calculating the absolute distances, independent of node features. Lastly, since we aim to identify high-risk network connections 
, we embed the shortest path distance as an edge feature.

\noindent\textbf{Anchor Sets.} We formulate a strategy for selecting anchors and assigning critical assets to anchor sets. Let $\mathit{n}$ be the number of highly-critical assets in the network. We first mark anchors around the nodes representing highly-critical assets where each anchor set holds only one critical asset. As per the original P-GNN model, to guarantee low distortion embedding at least $\mathit{k}$ anchors are sampled where $k=c \log ^2 |V|$ and $\mathit{c}$ is a constant. If the number of critical assets $\mathit{|V_{critical}|< k}$, the remaining anchors are sampled randomly where each node in $\mathit{V \sim V_{critical}}$ is sampled independently. The anchors' size is distributed exponentially and is calculated as follows:
\begin{equation}
|Anchor\_i|=\lfloor \frac{|V|}{2^{i+1}} \rfloor, i \in \{0..k \}
\end{equation}

\noindent\textbf{Objective Function. }The goal of the $SPGNN$ is to learn a mapping $\mathit{V \times V_{critical}^{k}} \mapsto  R^{+}$ to predict the actual minimum shortest path distances from each $\mathit{u \in V}$ to $\mathit{V_{critical}}$ where $\mathit{k = |V_{critical}|}$. Hence, unlike the original P-GNN objective function, defined for the downstream learning tasks using the learned positional embeddings (e.g. membership to the same community), our objective is formulated for learning the actual shortest path length as follows:
\begin{equation} \label{eq:loss}
\begin{aligned}
&\min_\phi \sum_{\forall u \in V} \mathcal{L}\left( \min_{i \in\{1\dots k\}}\hat{d}_\phi\left(u, v_{i}\right)-\min_{i \in\{1\dots k\}} d_{y}\left(u, v_{i}\right)\right)\\
&\min_\phi \sum_{\forall u \in V} \mathcal{L}\left(\min \left(\hat{d}_\phi\left(u, V_{critical}\right)\right)-\min \left(d_y(u, V_{critical}\right)\right)
\end{aligned}
\end{equation}
where $\mathcal{L}$ is the mean squared error (MSE) loss function to be minimized. $\mathit{\hat{d}_\phi\left(u, V_{critical}\right)}$ is the vector of learned approximation of the shortest path distance from a node $\mathit{u}$ to every critical asset $\mathit{v} \in V_{critical}$. $\mathit{d_{y}}$ is the observed shortest path distance. As the model aims to identify the risk imposed by critical paths, we account for the worst-case scenario by considering the minimum shortest path length from the (vulnerable) node to a highly-critical asset. Therefore, the loss is computed only on the minimum of the distance vector.

\noindent\textbf{Message Passing. }The message-passing function, in our approach, exclusively relies on the position information to calculate the absolute distances to the anchor sets and disregards node features. To calculate position-based embeddings, we follow the original P-GNN $q$-hop approach where the 1-hop $\mathit{d_{sp}^1}$ distance can be directly inferred from the adjacency matrix. During the training process, the shortest path distances $\mathit{d_{sp}^q(u,v)}$ between a node $u$ and an anchor node $v$ are calculated as follows~\cite{PGNN}:

\begin{equation}\label{eq:messagePassing}
\begin{aligned}
d_{sp}^q(u,v) \mapsto   \begin{cases} 
      d_{sp}(u,v), & if\; d_{sp}(u,v) < q \\
      \infty & otherwise. 
   \end{cases}
\end{aligned}
\end{equation}
Where $\mathit{d_{sp}(u,v)}$ is the shortest path distance between a pair of nodes. Since the P-GNN aims to map nodes that are close (in position) in the network to similar embedding, the distance is further mapped to a range in $\mathit{( 0,1)}$ as follows~\cite{PGNN}:
\begin{equation}\label{eq:message}
s(u,v) = \frac{1}{d_{sp}^q(u,v)+1}
\end{equation}
Accordingly, the message-passing process is defined as: 
\begin{equation}\label{eq:positionEmbedding}
    h_u=\phi(x_u \oplus_{(v \in \aleph_v)}\psi(u,v))
\end{equation}
where $\mathit{h_u}$ represents the node embedding of the vertex $\mathit{u}$, $\mathit{x_u}$ is the input feature vector of the node $\mathit{u}$ inferred based on the adjacency matrix. $\mathit{\oplus}$ is the aggregation function. In our approach, we found that the mean aggregation function provides the best performance. $\mathit{\psi}$ is the message function and is computed as described in Equation~\ref{eq:message}. Finally, $\mathit{\phi}$ is the update function to obtain the final representation of node $\mathit{u}$.

\noindent\textbf{Recovery of true paths length.} We aim to learn the true shortest path length by pulling the value of the node embedding closer to the labels during the learning process. To this end, we rely on the MSE loss function to minimize the deviation between the predicted and observed shortest path distances. To recover the true path length from the learned positional embedding, we introduce four steps to the P-GNN learning process after learning the node embeddings through message passing: Firstly, for every node $\mathit{u\in V}$, we calculate the absolute distance (AD) of the learned node embeddings between $\mathit{u}$ and each critical asset $\mathit{v \in V_{critical}}$. Secondly, we assign the minimum value of the calculated AD to the node $\mathit{u}$. Thirdly, as the calculated AD is not necessarily an integer, we approximate the assigned AD to an integer value representing the predicted shortest path distance. Lastly, we attribute the approximated shortest path value to the incident edge features.

\noindent\textit{(1) \underline{Absolute Distance (AD) of node embedding}.} We particularly use the AD function since it is less impacted by outliers and, hence, more robust. This is particularly significant since complex network structures are characterized by a high variance in the criticality of the assets and the path-length distributions. For every node $\mathit{u \in V}$, we calculate a vector of absolute distances $\mathit{T_u}$ between the learned embedding of $\mathit{u}$ denoted as $\mathit{h_u}$ and the embedding of every critical asset $\mathit{v_i \in V_{critical}}$, denoted as $\mathit{h_{v_i}}$. $\mathit{h_u}$ and $\mathit{h_{v_i}}$ are calculated as described in Equation~\ref{eq:positionEmbedding}. The AD vector is calculated as follows, where $\mathit{k}$ is the embedding space dimension:
\begin{equation}
\begin{aligned}
AD (u,v) = \sum_{n=1}^{k}|h_u^{n} - h_v^{n}|\\
T_u = \forall_{v_i \in V_{critical}} AD(u,v_i)
\end{aligned}
\end{equation}
$\mathit{T_u}$ is then used in Equation~\ref{eq:loss} to calculate the loss where $\mathit{\hat{d}\left(u, V_{critical}\right)=T_u}$.

\noindent\textit{(2) \underline{Minimum absolute distance to a critical asset}.} The downstream task is concerned with identifying the risk imposed by potential attack paths. If a node $\mathit{u \in V}$ has (shortest) paths to multiple critical assets, we account for the worst-case scenario by identifying the minimum length of the shortest paths $\mathit{z_u}$ and assigning its value as a feature for node $\mathit{u}$, as follows: 
\begin{equation}\label{eq:minimumSP}
z_u= \min_{i \in\{1\dots k\}} T_u^i
\end{equation}
where $\mathit{k}$ is the embedding space dimension.

\noindent\textit{(3) \underline{Approximation of path length}.} We identify two approaches for approximating the learned minimum shortest path length $\mathit{z_u}$ of a certain node $\mathit{u}$. The first approach, denoted as $SPGNN_R$, relies on simple rounding of the shortest path length. This naive approach is rather intuitive and is fully transferable as discussed in Section~\ref{sec:evaluation}. The predicted distance $\mathit{SP_{R}(u)}$ is then calculated as follows: 
\begin{equation}
\begin{aligned}
SP_{R}: V \mapsto N\\
SP_{R}(u) \mapsto Round(z_u)
\end{aligned}
\end{equation}

The second approach, $SPGNN_{DNN}$, relies on a deep neural network (DNN) to learn a mapping between the learned shortest path length and its integer representation. To overcome the inaccuracies induced by rounding the AD, we aim to increase the separation between the labels representing the observed path length. Since the downstream task is concerned with assessing the risks imposed by the attack paths, we restrict the detection of paths to a certain range of length values that are anticipated to induce high risks. Accordingly, we transform the path identification into a classification task where the learned embeddings are mapped to a class representing a path length within the range of interest.

The goal of the DNN is to learn a mapping to predict the integer representation of the minimum shortest path distance $\mathit{z_u}$ described in Equation~\ref{eq:minimumSP} from each $\mathit{u \in V}$ to $\mathit{V_{critical}}$ where $\mathit{k = |V_{critical}|}$. Accordingly, the objective function is:

\begin{equation} \label{eq:nn_Loss}
\min_{\theta}\sum_{\forall u \in V} \mathcal{L}_{c}(g_{\theta}(\lambda_{u}),l)
\end{equation}
where $g_{\theta}:R^{a}\mapsto R^{b}$ is a function that maps the node features $\mathit{\lambda_u}$ (that include $\mathit{z_u}$) where $\mathit{|\lambda_u| = a }$ to a label $\mathit{l}$ in the set of the encoded labels $\mathit{L= {1,...,b}}$ where $\mathit{b}$ is the threshold of paths length considered. $\mathit{\theta}$ denotes the parameters of $\mathit{g_\theta}$ and $\mathcal{L}_{c}$ is the categorical cross entropy loss function.

In addition to the minimum shortest path distance $\mathit{z_u}$, we enrich the classifier input with additional heuristics of the original P-GNN positional embeddings $\mathit{h_{u}}$ described in Equation~\ref{eq:positionEmbedding}. We rely on the intuition that the learned P-GNN embeddings of nodes that share the same shortest path distance are most likely to have similar statistical features. We define the DNN classifier input feature vector $\mathit{\lambda_u |\;\forall u \in V}$ as follows:
\begin{equation}
\begin{aligned}
\lambda_{u}= &(\max_{v \in V_{critical}}|cos_{sim}(u,v)|, \max_{v\in V_{critical}}cross_{entropy}(u,v),\\ &min(h_{u}), max(h_{u}), mean(h_{u}),var(h_{u}),norm_{2}(h_{u}),\\ &std(h_{u}), median(h_{u}), z_u).
\end{aligned}
\end{equation}

The output of the DNN model is the classes representing the different shortest path lengths. We rely on the one-hot encoding mapping to represent the output. The predicted distance denoted as $\mathit{SP_{DNN}(u)}$ is then calculated as follows: 
\begin{equation}
\begin{aligned}
SP_{DNN}: V \mapsto N\\
SP_{DNN}(u) \mapsto g_{\theta}(z_u)
\end{aligned}
\end{equation}

The stacking of a DNN classifier significantly enhances the accuracy of the $SPGNN$ when trained and tested on the same network data. However, it does not perform equally well in a transfer learning setting as discussed later in Section~\ref{sec:evaluation}. This can be attributed to the fact that the input to the DNN classifier depends on the learned positional embeddings $\mathit{h_u}$ and is highly impacted by the size and distribution of the anchors set.

\noindent\textit{(4) \underline{Shortest path as edge feature}.} When it comes to graph representation learning, relying on node features is often more efficient than edge features due to the amount of information contained within the nodes, and the relatively smaller number of nodes versus edges. As a result, we begin by predicting the shortest paths as node features. Then, we attribute the calculated distance to all {\it incident edges of the node}.
Let $\mathit{v}$ be a node in the network, $\mathit{SP(v)}$ be the learned integer representation of the minimum shortest path for $\mathit{v}$, and $\mathit{y_e}$ be the feature vector for edge $\mathit{e}$. Accordingly, the node features are assigned to their incident edges as follows:
\begin{equation}
 \{\forall u \in V \; \land \; \exists \;e_{u,v} \in E\;, \;y_{e_{u,v}} = SP(v)\} 
\end{equation}

\noindent\textbf{Labels. }Manually generated labels are expensive and hard to acquire. Therefore, we rely on a regular shortest path algorithm (e.g. Dijkstra), denoted by $\mathit{d_{sp}(u,v)}$, to generate the labels for training the $SPGNN$. We calculate the observed shortest path $\mathit{d_{y}}$ from a node $\mathit{u}$ to critical asset $\mathit{v}$ as per Equation~\ref{eq:osp}. The calculated shortest path represents the label of the node $\mathit{u}$.
\begin{equation}\label{eq:osp}
\begin{aligned}
d_{y}(u,v) \mapsto   \begin{cases} 
      0 & if\; v \notin V_{critical} \;\lor\; d_{sp}(u,v)=  \emptyset \\
      d_{sp}(u,v) & otherwise 
   \end{cases}
\end{aligned}
\end{equation}
 
\subsection{Risk Triage and Mitigation}\label{sec:criticality}

We develop a module to automate the assessment of risks imposed by potential exploitation of the detected attack paths in terms of the propensity and impact of compromising highly-critical assets. We first identify critical attack paths that require immediate intervention based on a pre-defined set of criteria. We then autonomously locate connections (edges) playing a key role in enabling the critical attack paths. Accordingly, we proactively implement the proper mitigation actions.

To assess attack path criticality, we introduce a new metric namely {\it Application Criticality (AC)}. The assets criticality metric assesses the risk based on the assets workload (e.g. database, application server, etc.) and data processed. However, the AC metric assesses the risk based on the application the asset belongs to. For instance, a human-resources application database with human-identifiable information is assigned a higher AC rating than an inventory application database.

\noindent\textbf{Application criticality: }Applications can be classified based on the scope of expected damages, if the application fails, as either, mission-critical, business-critical, or non-critical (operational and administrative)~\cite{app_criticality}. Organizations rely on mission-critical systems and devices for immediate operations. Even brief downtime of a mission-critical application can cause disruption and lead to negative immediate and long-term impacts. A business-critical application is needed for long-term operations and does not always cause an immediate disaster. Finally, organizations can continue normal operations for long periods without the non-critical application. Two different companies might use the same application but it might only be critical to one. Hence, we rely on the security team of enterprises contributing to this study to assign the AC.

Attack paths are considered critical if they meet the following criteria: (1) The start of the path is an asset with criticality level $\leq$ 4 implying the ease of accessibility of the asset. (2) Destination highly-critical assets belong to a mission-critical application (3) The shortest path is of length at most five.

After filtering out non-critical paths, we aim to locate and characterize connections playing a key role in enabling critical attack paths. Accordingly, we model a DNN edge classifier to assess the edges of attack paths. Three output classes are defined, based on which mitigation actions are planned: (1) Non-compliant critical, (2) compliant critical, and (3) safe.

Non-compliant edges are inherently un-trusted as they do not comply with the organization's communication best practices. Accordingly, non-compliant critical edges are immediately blocked by automatically tuning the ZT policies enabling the connection. Compliant connections represent legitimate organizational communication, hence blocking them might disrupt the network functionalities. Therefore, these connections are highlighted and the associated ZT policies are located. A system warning is generated requesting the network administrators to further assess the highlighted ZT policies. Finally, no actions are required for safe connections.

We assess the criticality of attack paths' edges based on the following criteria, representing the input of the DNN classifier: 
\begin{itemize}[noitemsep,topsep=0pt]
    \item Feature$_1$: The trust level of the edge destination asset.
    \item Feature$_2$: The AC rating of the edge destination asset.
    \item Feature$_3$: Exploited vulnerability base score of the source asset.
    \item Feature$_4$: The shortest path distance from the edge to the highly-critical asset.
     \item Feature$_5$: The compliance of the edge.
\end{itemize}

Let $f_\psi: E \mapsto \mathit{Y}$ be a function that maps the set of edges $\mathit{E}$ to the set of labels $\mathit{Y}$ representing the three edge classes where $\mathit{\psi}$ denotes the parameters of $\mathit{f_\psi}$. Let $feat_{e}$ be the input feature vector of the edge $\mathit{e}$ to be assessed. To optimize the edge's classification task, we express the objective function as the minimization of the cross-entropy loss function $\mathcal{L}_{d}$. 
We represent this objective function as follows:
\begin{equation} \label{eq:nn_Loss}
\min_{\psi}\sum_{\forall e in E } (f_{\psi}(feat_{e}),y)
\end{equation}

\section{Results and Evaluation}
\label{sec:evaluation}
The evaluation process is three folds: (1) evaluating the performance of the $SPGNN$ shortest path calculation in a semi-supervised setting (Sec.~\ref{sec:evsp}), (2) assessing the performance in a transfer-learning setting (Sec.~\ref{sec:transfer}), and (3) evaluating the accuracy of identifying critical attack paths and locating key edges (Sec.~\ref{sec:attack}).

\subsection{Experimental Settings} \label{sec:setup}
We test the performance of $SPGNN$ in three settings: 

\noindent{\bf Experiment 1 -- evaluating the performance of shortest paths identification. }
This experiment evaluates the ability of $SPGNN_R$ and $SPGNN_{DNN}$ to identify the shortest paths in a semi-supervised setting. We use the same dataset for training and testing. We compare the prediction accuracy with the baseline model $SPAGAN$. To identify the minimum ratio of labeled data required to achieve satisfactory performance, we use the train and test split masks with distribution shifts for all datasets described in Section~\ref{sec:data}. 

\noindent{\bf Experiment 2 -- assessing and validating the learning transferability.}
This experiment setting is particularly concerned with assessing the learning transferability of the proposed $SPGNN_R$ shortest path identification. We test the transferability by training the model using a dataset and testing it using a different unlabeled dataset. 

\noindent{\bf Experiment 3 -- Assessing and validating the attack paths identification.}
This experiment aims to assess the end-to-end performance of the $\mathit{SPGNN-API}$ in identifying critical attack paths and highlighting key connections enabling the paths. We test the performance of this task by comparing the model accuracy to labeled synthetic network datasets and real-world datasets of enterprises contributing to this research.

\subsection{Dataset} \label{sec:data}

Two classes of datasets are used for the proposed model evaluation: (1) Enterprise network datasets (two synthetic datasets, $STD_1$ and $STD_2$, and two real-world datasets, $RTD_1$ and $RTD_2$). (2) Two widely used citation network datasets Cora\cite{cora} and Citeseer~\cite{CiteSeer}. 

We generate two synthetic datasets ($STD_1$ and $STD_2$) to imitate a mid-sized enterprise network setting. We defined the node configurations and network connections to cover all possible combinations of values for the five features used for assessing the criticality of the attack path's edges.

We collect the real-world datasets, denoted by $RTD_1$ and $RTD_2$, from two-mid sized enterprises; a law firm and a university, respectively. We rely on the Nessus scan output to identify the configurations and properties of the network assets as well as the underlying vulnerabilities. We use enterprise-provided firewall rules, ZT policies, and governance rules to define and characterize the assets' communications. 

Table~\ref{tab:Data_stat} lists the details of the datasets used in assessing the performance of our proposed model. In the proposed approach, we identify the path length to a set of anchor nodes to represent highly-critical assets. For the citation datasets, we randomly sample nodes to represent highly-critical assets. Since the citation datasets do not represent a real network setting, we will limit the evaluation of the attack path identification to the (real-world and synthetic) enterprise network datasets. 

\begin{table}[h]
\tabcolsep=0.12cm
\begin{tabular}{l||r|r|r|r|r}
\hline
\toprule 
\textbf{Dataset} & \textbf{Nodes}& \textbf{Edges} & \makecell{\textbf{Critical} \\ \textbf{Assets}} & \makecell{\textbf{Compliant} \\ \textbf{Edges}} & \makecell{\textbf{Non-compliant} \\ \textbf{Edges}}\\
\hline
$SDT_1$ & 864& 5,018 & 284 &2,002&3,016\\
$SDT_2$ & 865& 5,023 & 284 &2,002&3,021\\
$RTD_1$ & 221& 1.914 & 21 &882&1,032\\
$RTD_2$ & 370& 21,802& 70 &10901&10901\\
$CORA$ & 2.708& 10.556 & 180 &N/A&N/A\\
$CITESEER$ & 3.327& 9.464 & 523 &N/A&N/A\\
\hline\bottomrule       
\end{tabular}
    \caption{\small Dataset features and statistics.} 
    \label{tab:Data_stat}
\end{table}

\subsection{Baseline Models}
We compare the performance of our proposed model architectures $SPGNN_R$ and $SPGNN_{DNN}$ with the state-of-the-art baseline $SPAGAN$~\cite{Spagan} with respect to the shortest path identification. The $SPAGAN$ conducts path-based attention that explicitly accounts for the influence of a sequence of nodes yielding the minimum cost, or shortest path, between the center node and its higher-order neighbors. 
 
To validate the performance of the $\mathit{SPGNN-API}$ for attack paths identification, we generate the network attack graph using the $\mathit{MulVAL}$ tool~\cite{mulval2} by combining the output of the vulnerability scanner Nessus~\cite{nessus} and the enterprise network perimeter and ZT firewall policies. 

\subsection{Evaluation of Shortest path Detection}\label{sec:evsp}

In this section, we assess the performance of the two proposed architectures the $\mathit{SPGNN_R}$ and $\mathit{SPGNN_{DNN}}$ using all six datasets. We report the mean accuracy of 100 runs with 80\%-20\% train-test masks and 20 epochs.

\noindent\textbf{Accuracy evaluation: }Table~\ref{tab:part1_eval} summarizes the performance of $\mathit{SPGNN_R}$ and $\mathit{SPGNN_{DNN}}$. While both models can discriminate symmetric nodes by their different distances to anchor sets, we observe that $\mathit{SPGNN_{DNN}}$ significantly outperforms $\mathit{SPGNN_R}$ across all datasets. This can be attributed to the power of the DNN in capturing the skewed relationships between the generated positional embedding and the defined set of path-length classes. Furthermore, transforming the prediction of path lengths to a classification task, where one-hot encoding is used to represent the output, enables the model to capture the ordinal relationships between the different lengths and hence the gain in the performance. Both architectures exhibit performance degradation when tested with the real-world dataset $RTD_1$. Due to the relatively small size of the dataset. The model could not capture the complex relationships between the network entities during training. 

\noindent\textbf{Self-loops: }In general, adding self-loops allows the GNN to aggregate the source node's features along with that of its neighbors~\cite{self-loops}. Nevertheless, since our model relies only on positional embedding irrelevant to the features of the nodes, removing the self-loops enhances the accuracy of $SPGNN$ as detailed in Table~\ref{tab:part1_eval} as the iterative accumulation of the node positional embedding confuses the learned relative distance to the anchor sets. Accordingly, we introduce a data pre-processing stage to remove self-loops in the real-world network datasets and the citation datasets.

\noindent\textbf{SPGNN convergence: }We illustrate in Figure~\ref{Loss} the progression of the Mean Squared Error MSE loss during the training process of $\mathit{SPGNN_R}$. We particularly assess the $\mathit{SPGNN_R}$ since, unlike the $\mathit{SPGNN_{DNN}}$, its output directly reflects the GNN performance without further learning tasks. We observe that the gradient is sufficiently large and proceeding in the direction with the steepest descent which indeed minimizes the objective. The stability and efficacy of the learning process constantly enhance the accuracy of the model irrelevant of the dataset characteristics. The objective function is sufficiently smooth indicating that the model is not under-fitting.  

\begin{figure}
    \centering
    \includegraphics[width=0.48\textwidth, height=0.18\textheight]{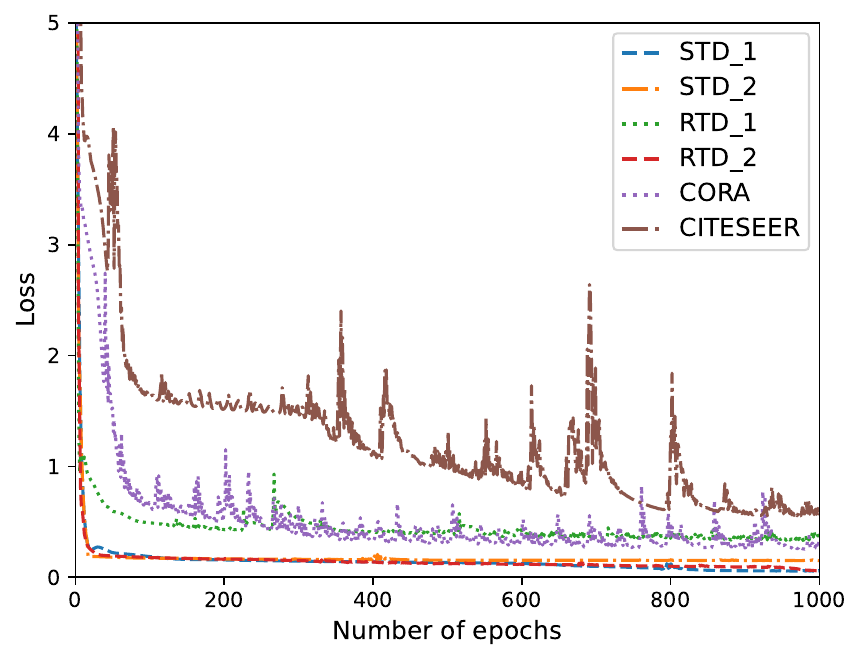}
    \caption{\small $SPGNN_R$ MSE loss convergence for the six datasets.} 
    \label{Loss}
\end{figure}

\noindent{\bf Analysis of the SPGNN$_\textbf{R}$ generated shortest path distance embedding.} We conducted an in-depth analysis of 20 random samples from the test sets of the six datasets. For each sample, we plot the predicted $\mathit{\hat{d}}$ vs rounded $SP_{pred}$ vs observed $\mathit{d_y}$ shortest paths distances in blue, yellow, and red, respectively, as illustrated in Figure~\ref{fig:SP_investigation}. We observe the proximity of the predicted and observed distances where the predicted values are mostly in the range of +/- 1 hop from the observed values. Hence, we prove the strength of the proposed GNN approach in approximating the shortest path distance. We further notice that the rounded values are vastly overlapping with the observed values which further proves the robustness of the simple, yet intuitive, rounding approach. 

\begin{figure*}
    \centering
     \vspace{-0.25cm}
    \subfloat[STD$_{1}$]{\includegraphics[width=0.33\textwidth, height=0.15\textheight]{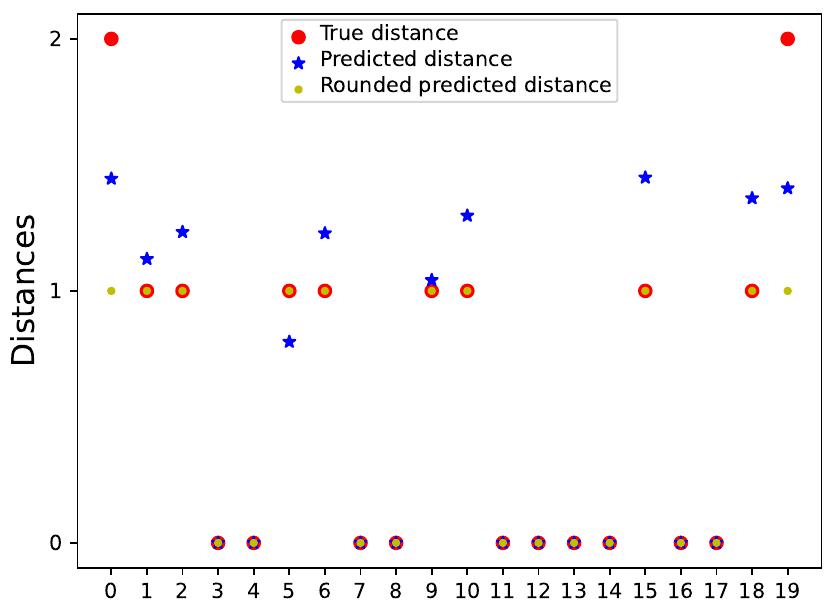}}
     \subfloat[STD$_{2}$]{\includegraphics[width=0.33\textwidth, height=0.15\textheight]{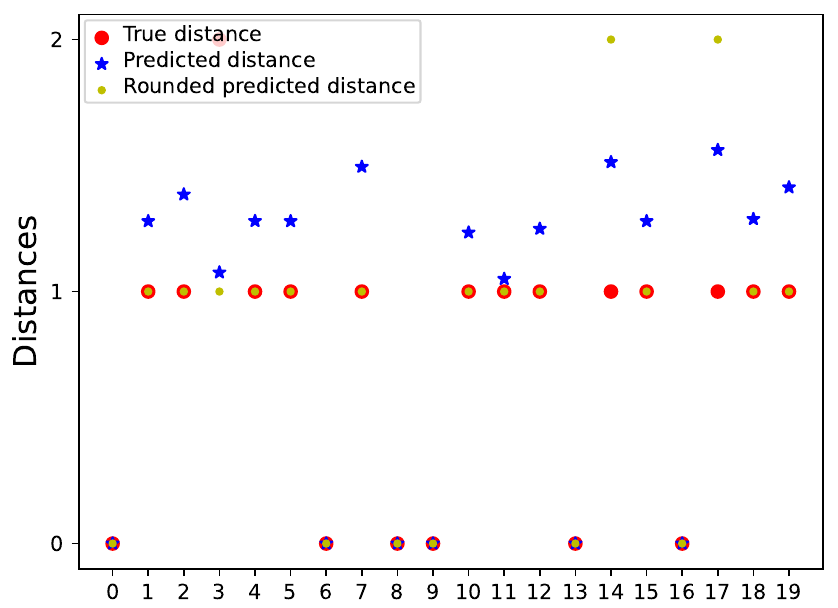}}
     \subfloat[RTD$_{1}$]{\includegraphics[width=0.33\textwidth, height=0.15\textheight]{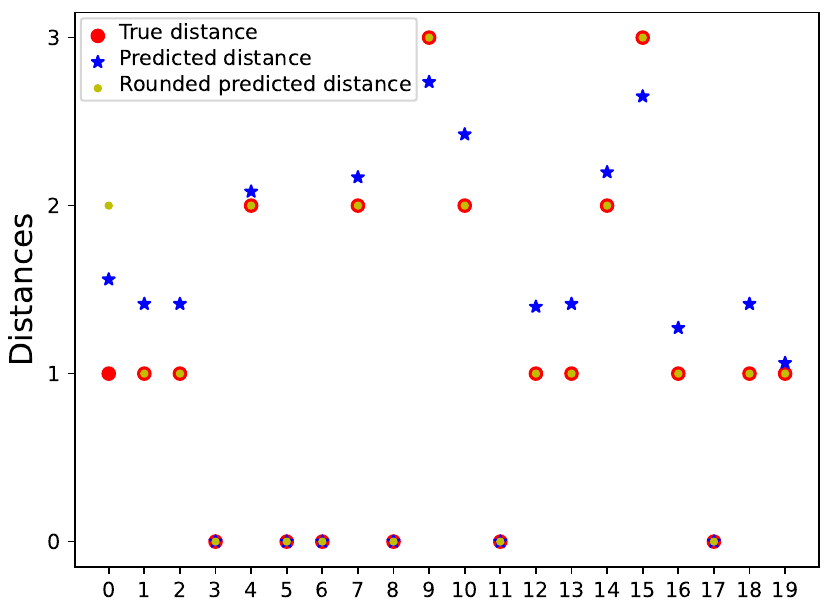}} \\
     \subfloat[RTD$_{2}$]{\includegraphics[width=0.33\textwidth, height=0.15\textheight]{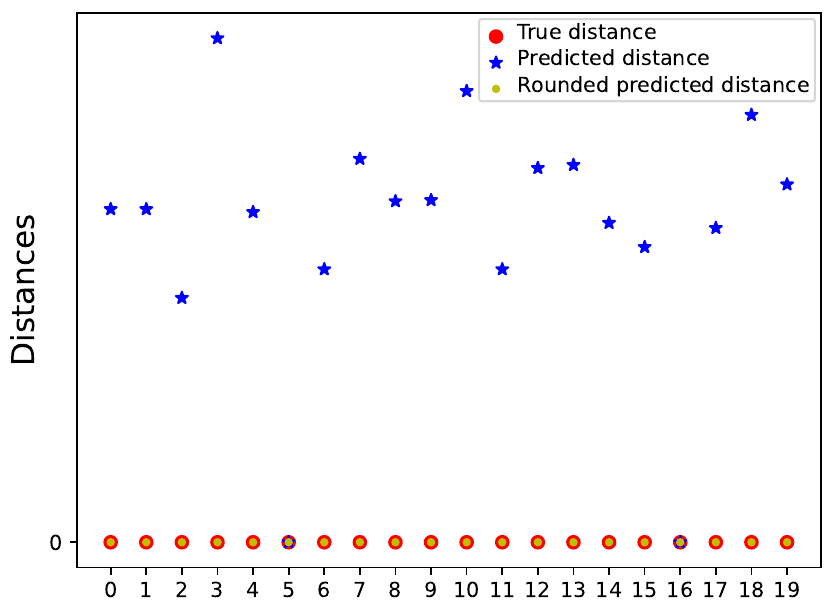}}
     \subfloat[Cora]{\includegraphics[width=0.33\textwidth, height=0.15\textheight]{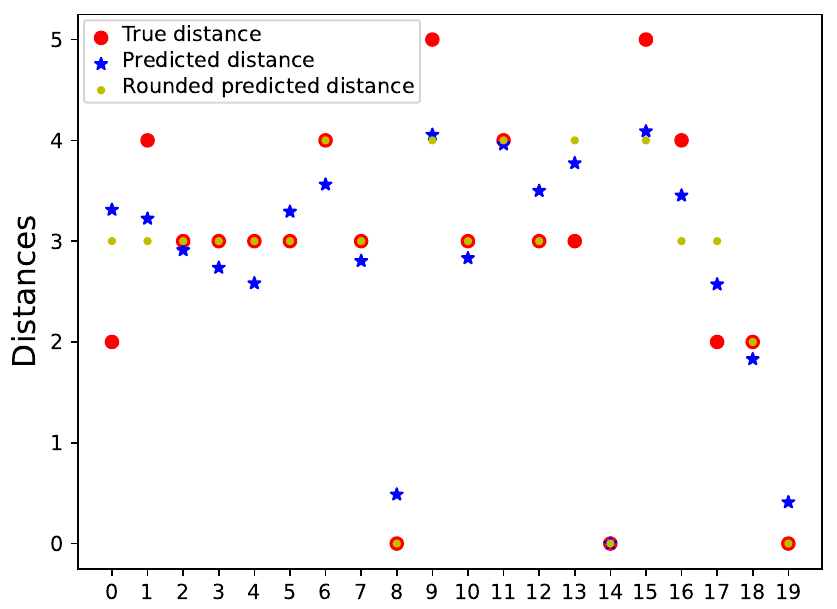}}
     \subfloat[Citeseer]{\includegraphics[width=0.33\textwidth, height=0.15\textheight]{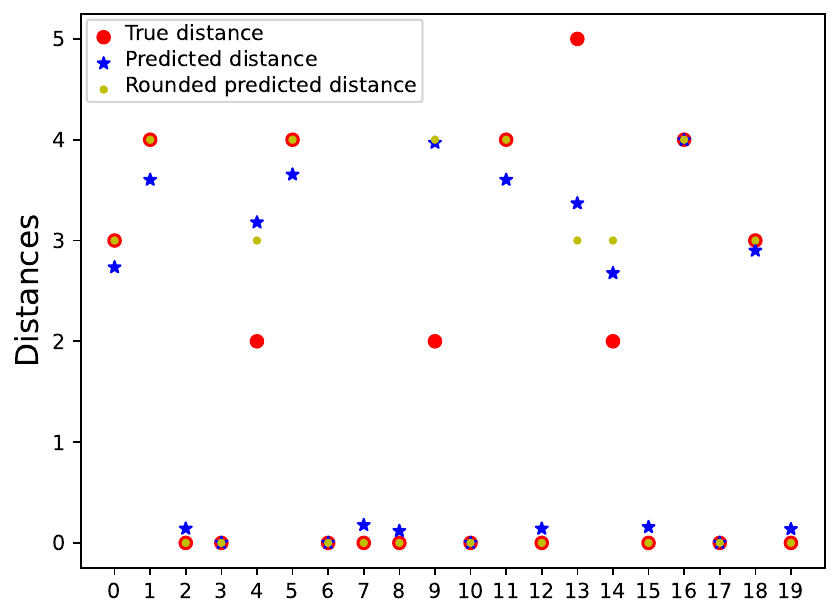}}
      \vspace{-0.20cm}
    \caption{\small Shortest path distance distribution of 20 random samples from each of the six datasets. The blue and yellow points are the $SPGNN$ predicted distances {\it before} and {\it after} the application of the rounding process, respectively. The red points are the observed distances. The Figures illustrate the accuracy of the predicted distances being within the range [-1,1] of the observed values. We further observe that the majority of the {\it rounded distances} are either overlapping with or closer to the {\it observed distances}. This shows the efficiency of the rounding approach to enhance the shortest path distance prediction accuracy.} 
    \label{fig:SP_investigation}
    \vspace{-0.25cm}
\end{figure*}

\begin{table*}[h]
\centering
\tabcolsep=0.09cm
\begin{tabular}{lcccccc|cccccc}
\hline
\toprule 
         & \multicolumn{6}{l}{\bf  Dataset Before Deleting self loops}& \multicolumn{6}{l}{\bf  Dataset After Deleting self loops} \\
         \cline{2-3}\cline{4-7} \cline{8-13}
{\bf Metrics}   & $STD_1$          & $STD_2$  & $RTD_1$  & $RTD_2$       & $CORA$         & $CITESEER$  & $STD_1$          & $STD_2$  & $RTD_1$  & $RTD_2$       & $CORA$         & $CITESEER$ 
\\ \hline
$SPGNN_R$ $\mathcal{L}$ & $0.07$             & $0.14$ & $0.36$ & $0.02$          & $0.33$             & $0.53$   & $0.02$             & $0.14$ & $0.22$ & $0.01$          & $0.29$             & $0.38$  \\
\hline
Accuracy $\mathit{SP_{pred}}$       & $90.00\%$             & $84.04\%$ & $71.00\%$ & $94.02\%$           & $65.70\%$             & $68.53\%$   & $98.88\%$             & $84.47\%$ & $72.41\%$ & $97.05\%$           & $65.34\%$             & $72.53\%$ \\\hline
$Accuracy_{\pm {1}hop}$ & $100\%$             & $98.42\%$      & $91.61\%$     
& $100\%$ & $96.41\%$             & $92.65\%$ & $100\%$             & $98.50\%$      & $93.85\%$     
& $100\%$ & $97.11\%$             & $94.54\%$\\
\hline
\hline
$SPGNN_{DNN} \mathcal{L}_{c}$    & $0.03$             & $0.08$ & $0.24$  &$0.01$         & $0.26$             & $0.41$ & $0.01$             & $0.10$ & $0.19$  &$0.01$         & $0.23$             & $0.26$ \\
\hline
$Accuracy \mathit{SPGNN_{DNN}}$    & $95.63\%$             & $80.41\%$ & $53.50\%$  &$96.10\%$         & $81.36\%$             & $79.36\%$    & $98.45\%$             & $84.74\%$ & $78.65\%$  &$98.52\%$         & $75.82\%$             & $81.20\%$  \\\hline
$Accuracy_{\pm {1}hop}$ & $86.45\%$             & $85.29\%$      & $86.15\%$     & $98.65\%$ &$92.70\%$             & $84\%$ & $93.10\%$             & $91.93\%$      & $89.23\%$     & $100\%$ &$92.94\%$             & $87.32\%$ \\ 

\hline
\hline
MSE(SPAGAN)     & $0.54$             & $0.62$ & $0.91$  &$0.48$         & $0.85$             & $0.95$  & $0.52$             & $0.59$ & $0.72$  &$0.35$         & $0.69$             & $0.82$ \\
\hline
$Accuracy \mathit{SP_{pred}}$     & $52.36\%$             & $50.14\%$ & $57.50\%$ &$82.35\%$           & $62.12\%$            & $53.36\%$  & $54.23\%$             & $52.36\%$ & $56.23\%$ &$85.65\%$           & $63.26\%$            & $55.68\%$ \\\hline
$Accuracy_{\pm {1}hop}$ & $86.45\%$             & $85.29\%$      & $86.15\%$     & $98.65\%$ &$92.70\%$             & $84\%$  & $88.20\%$             & $85.60\%$      & $84.42\%$     & $96.75\%$ &$93.98\%$             & $83.62\%$ \\ \hline\bottomrule       
\end{tabular}
     \vspace{-0.25cm}
    \caption{\small Overview of shortest paths identification accuracy of {\bf $SPGNN_R$} and {$SPGNN_{DNN}$} as compared to the $SPAGAN$ across the six datasets before and after deleting self-loops.}
    \label{tab:part1_eval}
     \vspace{-0.35cm}
\end{table*}

\noindent\textbf{Baseline comparison: }We compare the performance of the proposed model with the baseline $SPAGAN$. We observe that the proposed architectures, in particular the $SPGNN_{DNN}$, strictly outperform $SPAGAN$ and can capture the skewed relationships in the datasets as shown in Table~\ref{tab:part1_eval}. This can be attributed to the fact that $SPAGAN$ uses a spatial attention mechanism that only considers the neighboring nodes within a predefined radius around each target node during the learning phase and does not incorporate features of nodes beyond the predefined distance which impacts the model performance. Furthermore, $SPAGAN$ (and most state-of-the-art approaches) relies on the graph elements' features to calculate the shortest paths distance information. This justifies the performance degradation of $SPAGAN$, in this case, since only graph structure and positional embedding are considered. This further proves the strength of the proposed approach that can identify, with high accuracy, the shortest paths distance irrelevant to graph elements features.

\subsection{Evaluation of Transfer-Learning} \label{sec:transfer}
In this setting, the pre-training and testing processes are executed through distinct datasets. The goal of the pre-training is to transfer knowledge learned from labeled datasets to facilitate the downstream tasks with the unlabeled datasets. We only consider the $SPGNN_R$ for testing in this setting. The stacked DNN of the $SPGNN_{DNN}$ approach is characterized by a fixed input size and hence is not expandable to accommodate different network structures.

To assess the robustness of the model transferability, we pre-train the model using different synthetic and real-world datasets. We observe that, in general, the size and sophistication of the dataset used for pre-training highly impact the performance of the model transferability. In general, training with real data yields better performance. We believe that the significant improvements can be attributed to the ability of $SPGNN$ to utilize the perturbation in real-world data to consider more complicated interactions between the data samples which optimizes the model's ability to extend label information to unlabeled datasets. 

In contrast, pre-training the model with synthetic data and testing on a real dataset slightly hurts the accuracy. The non-perturbed structure of the synthetic data gives limited performance gain and yields negative transfer on the downstream classification task. In general, the results show convincing evidence that the inductive capabilities of the proposed $SPGNN$ generalize to unseen datasets as detailed in Table~\ref{tab:transferability evaluation}.

\begin{table}[h]
\tabcolsep=0.07cm
\centering
\begin{tabular}{lccc|ccc}
\hline
\toprule
         & \multicolumn{3}{l}{\bf Model trained by $SDT_1$} & \multicolumn{3}{l}{\bf Model trained by $RTD_1$} \\
         \cline{2-3}\cline{4-7}
{\bf Metrics}   & $SDT_2$        & $RTD_1$ &$RTD_2$         & $SDT_1$        & $STD_2$&$RTD_2$       
\\ \hline
$Accuracy \mathit{SP_{pred}}$ & $80.53\%$             & $66.51\%$    &$80.41\%$       &  $81.23\%$             & $79.45\%$  &$66.02\%$               \\
\hline
$Accuracy_{\pm {1}hop}$     &  $99.65\%$             & $95.65\%$   &$97.08\%$          &  $97.10\%$             & $95.64\%$      &$95.45\%$         \\
\hline\bottomrule       
\end{tabular}
    \caption{\small Transfer-learning evaluation of $SPGNN_R$.}
    \label{tab:transferability evaluation}
\end{table}

\subsection{Evaluation of Attack Paths and Critical Edges Detection} \label{sec:attack}
The $\mathit{SPGNN-API}$ does not record the end-to-end sequence of attack steps as there might be an infinite number of alternatives as discussed in Section~\ref{sec:attack}. It rather identifies the propensity of an edge being part of an attack, i.e. there exists a (shortest) path from that edge to a highly-critical asset going through vulnerable nodes. Accordingly, to evaluate the performance of the attack path detection, we do not rely on an end-to-end assessment of attack paths. We rather assess the propensity of single edges being part of an attack. We evaluate the accuracy of the edge classification (Sec.~\ref{sec:criticality}) in two different settings semi-supervised and transfer-learning. We compare the model performance against a baseline $\mathit{MulVAL}$. 

We base our assessment on the four enterprise network datasets as the citation datasets do not incorporate vulnerability information. We rely on the security team of the enterprises contributing to this study to manually label connections they would potentially block or patch given the network structure, reported vulnerabilities, and network visualization tools. 

\noindent\textbf{Accuracy assessment: }We assess the performance of the edge classifier in categorizing the attack path edges as either critical compliant, critical non-compliant, or safe. Comparing the output of the classifier to the manually labeled data we note the performance results in Table~\ref{tab:attack_path_accuracy}. Since the set of safe edges comprises the attack path edges classified as safe as well as the connectivity graph edges that were not part of any attack path, the recorded model accuracy proves the efficacy of the presented approach in detecting attack paths in general and identifying key critical edges in particular.

In addition to the raw accuracy rates, we report the receiver operating characteristic curve (ROC) and area under the curve (AUC). We assess the ability of the classifier to discriminate critical and safe edges, in general. Accordingly, we combine the critical compliant and critical non-compliant classes. The true positive samples are (compliant/non-compliant) critical samples that have been classified as critical. The false positive samples are critical samples that have been classified as safe. The ROC curve in Figure~\ref{fig:ROCTrain} illustrates outstanding discrimination of the two classes with an AUC score of 0.998. 

\begin{table}[h]
\centering
\tabcolsep=0.3cm
\begin{tabular}{lcccc}
\hline
\toprule 
         & \multicolumn{4}{l} {\bf  Dataset} \\
         \cline{2-3}\cline{4-5} 
{\bf Metrics}   & $STD_1$          & $STD_2$  & $RTD_1$ & $RTD_2$       
\\
\hline
Cross\_Entropy Loss & $0.095$             & $ 0.0061$ & $0.01$           & $0.007$     \\
\hline
Accuracy    & $99.5\%$             & $100\%$ & $ 99.11\%$           & $100\%$  \\

\hline\bottomrule     
\end{tabular}
    \caption{\small Performance overview of the $SPGNN$ edge criticality classification in semi-supervised setting.}
    \label{tab:attack_path_accuracy}
\end{table}

\begin{figure}[h]
    \centering
     \vspace{-0.20cm}
    \includegraphics[width=0.5\textwidth]{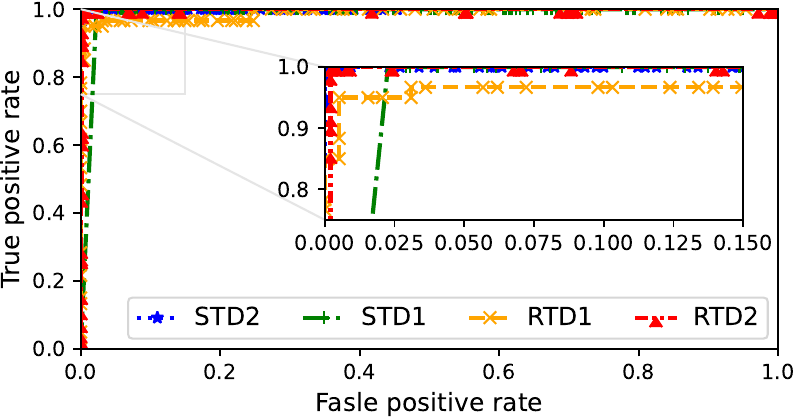}
     \vspace{-0.20cm}
    \caption{\small ROC curves of the $SPGNN$ edge classification in the semi-supervised setting.} 
    \label{fig:ROCTrain}
         \vspace{-0.25cm}
\end{figure}

\noindent\textbf{Transfer-learning: }To assess the end-to-end transferability of the presented approach, we train the edge classifier using a dataset and test it using different datasets. The recorded classification accuracy in Table~\ref{tab:tarnsfer_class} proves the inductive capabilities of $SPGNN$ and its ability to efficiently characterize previously unseen data. To our expectations, training the model using a real dataset performs better on all datasets. The model's capacity to extend the label information to previously unseen datasets is enhanced by the perturbations in real-world datasets that enable the classifier to consider more complex interactions between the data samples. To plot the ROC curve, we combine the critical compliant and critical non-compliant classes and assess the model's ability to discriminate the critical and safe edges. The ROC curve in Figure~\ref{fig:ROCTransfer} illustrates outstanding discrimination of the two classes with an AUC score between 0.93 and 0.98.

\begin{table}[h]
\tabcolsep=0.04cm
\begin{tabular}{lccc|ccc}
\hline
\toprule  
        & \multicolumn{3}{l}{\bf Model trained by $RTD_1$} & 
          \multicolumn{3}{l}{\bf Model trained by $STD_1$} \\
         \cline{2-3}\cline{4-7}
{\bf Metrics}   & $STD_1$          & $STD_2$  & $RTD_2$         & $STD2$        & $RTD1$   & $RTD2$ 
\\ \hline
Cross\_Entropy Loss & $0.009$             & $0.0037$ & $1.20$           & $0.002$   & $0.79$     & $0.18$    \\
\hline
Accuracy    & $100.00\%$             & $98.17\%$ & $92.75\%$           & $99.87\%$       & $92.42\%$   & $97.44\%$        \\\hline

\bottomrule       
\end{tabular}
     \vspace{-0.20cm}
    \caption{\small Performance overview of the $SPGNN$ edge criticality classification in transfer-learning.}
    \label{tab:tarnsfer_class}
     \vspace{-0.25cm}
\end{table}

\begin{figure}[h]
    \centering
    \includegraphics[width=0.45\textwidth]{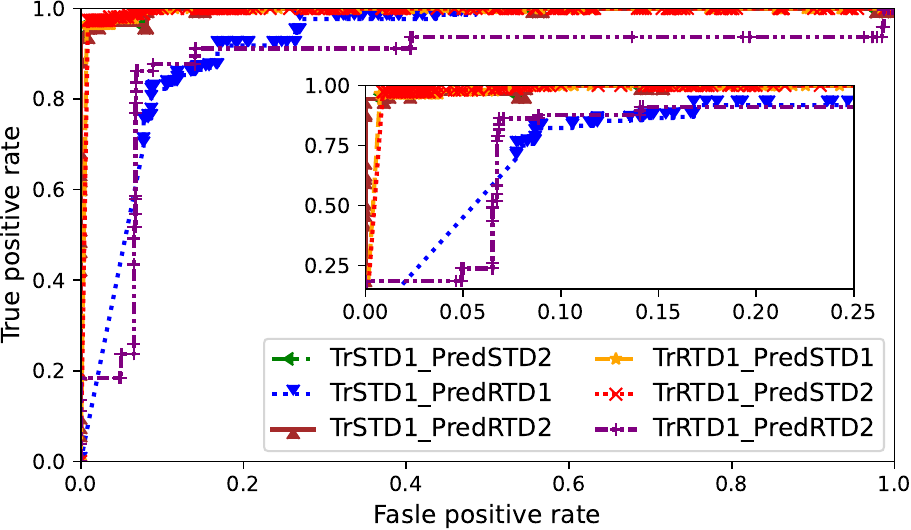}
    \caption{\small ROC curves of the $SPGNN$ edge classification in the transfer-learning setting.} 
    \label{fig:ROCTransfer}
    \vspace{-0.25cm}
\end{figure}

\noindent\textbf{Baseline comparison }: We compare the $\mathit{SPGNN-API}$ with the $\mathit{MulVAL}$ attack graph generator. The $\mathit{MulVAL}$ attack graph nodes can be of three types; configuration nodes, privilege nodes (exploits), and attack step nodes (conditions). The privilege nodes represent compromised assets. The root nodes of the attack graph represent network configurations/vulnerabilities contributing to attack possibilities. The privilege nodes denote the compromised assets. 

The set of paths of the attack graph comprises all directed attack paths starting at the root configuration nodes and ending at the privilege nodes (attack goals). We configure the attack path generation to have all highly-critical assets as attack goals. We assess the attack step nodes and note the ZT policies that have been a step to achieve the attack privilege. We then compare the noted rules to the set of rules that have been flagged as critical by the $\mathit{SPGNN-API}$.

\begin{table}[th]
\centering
\tabcolsep=0.2cm
\begin{tabular}{lcc|cc}
\hline
\toprule 
         & \multicolumn{2}{l} {\bf  SPGNN-API} &\multicolumn{2}{l} {\bf  MulVAL}\\
         \cline{2-3}\cline{4-5} 
{\bf State}   & $Detected$          & $Missed$  & $Detected$ & $Missed$       
\\
\hline
Critical Edges & $713$             & $ 5$ & $171$ & $542$     \\
\hline
Compromised Assets & $44$             & $ 0$ & $21$ & $25$ 
\\\hline
\bottomrule     
\end{tabular}
     \vspace{-0.20cm}
    \caption{\small Detected attack paths analysis of the $\mathit{SPGNN-API}$ as compared to the baseline $\mathit{MulVAL}$}
    \label{tab:part 2 evaluation}
    \vspace{-0.4cm}
\end{table}
We perform the analysis relying on $RTD_2$ since no Nessus scan is available for the synthetic datasets and we had limited access to the $RTD_1$ Nessus output for privacy reasons. The dataset has 370 nodes, of which 70 are highly-critical assets. The Nessus identified 44 vulnerable assets, of which six are highly critical. All six assets have been identified as potentially compromised by the $\mathit{MulVAL}$ as well as $\mathit{SPGNN-API}$. The $SPGNN$, however, outperformed the $\mathit{MulVAL}$ by detecting more potentially compromised non-critical assets as detailed in Table~\ref{tab:part 2 evaluation}.  This proves the significance of the presented holistic approach to vulnerability interaction.  

Further assessing the generated attack paths, we observe that $\mathit{SPGNN-API}$ labeled 713 edges (and underlying ZT policies) as critical while only 171 appeared as a $\mathit{MulVAL}$ attack step. This can be attributed to the fact that $\mathit{MulVAL}$ restricts the detection of potential attacks to a predefined set of vulnerability interactions while the $\mathit{SPGNN-API}$ assumes that any vulnerability can potentially be exploited.
Of the 171 edges detected by $\mathit{MulVAL}$, our approach detected 166. The missed five edges are connecting level 7 highly-critical assets to level 1 assets. Since we aim to protect highly-critical assets these edges are not considered critical as per our features.


\section{Conclusion}
\label{sec:conclusion}
This work presents the first attempt at GNN-based identification of attack paths in dynamic and complex network structures. Our work fills the gaps and extends the current literature with a novel GNN-based approach to automated vulnerability analysis, attack path identification, and risk assessment of underlying network connections that enable critical attack paths. We further present a framework for automated mitigation through a proactive non-person-based timely tuning of the network firewall rules and ZT policies to bolster cyber defenses before potential damage takes place.

We model a novel GNN architecture for calculating shortest path lengths exclusively relying on nodes' positional information irrelevant to graph elements' features. We prove that removing self-loops enhances the accuracy of shortest path distance identification as self-loops render the nodes' positional embedding misleading. Furthermore, our in-depth analysis of attack path identification proves the efficiency of the presented approach in locating key connections potentially contributing to attacks compromising network highly-critical assets, with high accuracy. A key strength of the presented approach is not limiting the attacks' detection to a predefined set of possible vulnerability interactions. Hence, it is capable of effectively and proactively mitigating cyber risks in complex and dynamic networks where new attack vectors and increasingly sophisticated threats are emerging every day.



 \balance
\bibliographystyle{IEEEtran}
\bibliography{bibl.bib}

\end{document}